\documentclass[aps, prd, nofootinbib]{revtex4}

\newcommand{\bb}{\textbf}
\newcommand{\bbb}{\bigskip}
\newcommand{\be}{\begin{eqnarray}}
\newcommand{\bee}{\begin{enumerate}}
\newcommand{\bit}{\begin{itemize}}

\def\bkt#1{\left(#1\right)} 
\def\bkts#1{\left[#1\right]} 
\def\bkta#1{\langle#1\rangle} 

\def\diff#1#2{{d{#1}\over d{#2}}}
\def\ee{\end{eqnarray}}
\newcommand{\eee}{\end{enumerate}}
\def\eg{\textit{e.g.} }
\newcommand{\eit}{\end{itemize}}
\def\etal{\textit{et al.}} 
\def\etals{{\it et al. }\rm} 
\def\etc{\textit{etc.} }
\def\fnl{f_{\mbox{\scriptsize NL}}}
\def\ff{\phantom{.}}

\def\gnl{g_{\mbox{\scriptsize NL}}}

\newcommand{\ii}{\textit}

\def\lab{\label}

\newcommand{\liner}{\centerline{\vrule height 0.01in width 4.4in}}

\newcommand{\mb}{\mathbf}
\def\mc#1{\mathcal{#1}}
\newcommand{\mmm}{\medskip}

\def\ng{non-Gaussianity }
\newcommand{\no}{\noindent}
\newcommand{\nn}{\nonumber}

\def\pr{\prime}
\def\re#1{(\ref{#1})}

\newcommand{\sss}{\smallskip}
\def\sub#1{_{\mbox{\scriptsize{#1}}}}
\def\sun{\odot}

\newcommand\aj{\ref@jnl{AJ}}%
\newcommand\actaa{\ref@jnl{Acta Astron.}}%
\newcommand\araa{\ref@jnl{ARA\&A}}%
\newcommand\apjl{\ref@jnl{ApJ}}%
\newcommand\apjs{\ref@jnl{ApJS}}%
\newcommand\apss{\ref@jnl{Ap\&SS}}%
\newcommand\aap{\ref@jnl{A\&A}}%
\newcommand\aapr{\ref@jnl{A\&A~Rev.}}%
\newcommand\aaps{\ref@jnl{A\&AS}}%
\newcommand\azh{\ref@jnl{AZh}}%
\newcommand\baas{\ref@jnl{BAAS}}%
\newcommand\caa{\ref@jnl{Chinese Astron. Astrophys.}}%
\newcommand\cjaa{\ref@jnl{Chinese J. Astron. Astrophys.}}%
\newcommand\icarus{\ref@jnl{Icarus}}%
\newcommand\jcap{\ref@jnl{J. Cosmology Astropart. Phys.}}%
\newcommand\jrasc{\ref@jnl{JRASC}}%
\newcommand\memras{\ref@jnl{MmRAS}}%
\newcommand\mnras{\ref@jnl{MNRAS}}%
\newcommand\na{\ref@jnl{New A}}%
\newcommand\nar{\ref@jnl{New A Rev.}}%
\newcommand\pasa{\ref@jnl{PASA}}%
\newcommand\pasp{\ref@jnl{PASP}}%
\newcommand\pasj{\ref@jnl{PASJ}}%
\newcommand\qjras{\ref@jnl{QJRAS}}%
\newcommand\rmxaa{\ref@jnl{Rev. Mexicana Astron. Astrofis.}}%
\newcommand\skytel{\ref@jnl{S\&T}}%
\newcommand\solphys{\ref@jnl{Sol.~Phys.}}%
\newcommand\sovast{\ref@jnl{Soviet~Ast.}}%
\newcommand\ssr{\ref@jnl{Space~Sci.~Rev.}}%
\newcommand\zap{\ref@jnl{ZAp}}%
\newcommand\iaucirc{\ref@jnl{IAU~Circ.}}%
\newcommand\aplett{\ref@jnl{Astrophys.~Lett.}}%
\newcommand\apspr{\ref@jnl{Astrophys.~Space~Phys.~Res.}}%
\newcommand\bain{\ref@jnl{Bull.~Astron.~Inst.~Netherlands}}%
\newcommand\fcp{\ref@jnl{Fund.~Cosmic~Phys.}}%
\newcommand\gca{\ref@jnl{Geochim.~Cosmochim.~Acta}}%
\newcommand\grl{\ref@jnl{Geophys.~Res.~Lett.}}%
\newcommand\jgr{\ref@jnl{J.~Geophys.~Res.}}%
\newcommand\jqsrt{\ref@jnl{J.~Quant.~Spec.~Radiat.~Transf.}}%
\newcommand\memsai{\ref@jnl{Mem.~Soc.~Astron.~Italiana}}%
\newcommand\nphysa{\ref@jnl{Nucl.~Phys.~A}}%
\newcommand\physrep{\ref@jnl{Phys.~Rep.}}%
\newcommand\physscr{\ref@jnl{Phys.~Scr}}%
\newcommand\planss{\ref@jnl{Planet.~Space~Sci.}}%
\newcommand\procspie{\ref@jnl{Proc.~SPIE}}%

\usepackage{amsmath}
\usepackage{amsfonts}
\usepackage{amssymb}
\usepackage{amsthm}
\usepackage{graphicx}
\usepackage{epsfig}
\usepackage{hyperref}
\usepackage[justification=raggedright,font=small]{caption}
\usepackage{subfig}

\usepackage{bm}

\begin{document}
\title{Scale-dependent Bias from the Reconstruction of Non-Gaussian Distributions}

\author{Sirichai Chongchitnan}
\author{Joseph Silk}
\affiliation{Department of Physics, University of Oxford and \\
Beecroft Institute for Particle Astrophysics and Cosmology,  \\ Denys Wilkinson Building, 1 Keble Road, Oxford, OX1 3RH, UK}
\email{siri@astro.ox.ac.uk}

\begin{abstract}
Primordial non-Gaussianity introduces a scale-dependent variation in the clustering of density peaks corresponding to rare objects. This variation, parametrized by the bias, is investigated on scales where a linear perturbation theory is sufficiently accurate. The bias is obtained directly in real space by comparing the one- and two-point probability distributions of density fluctuations. We show that these distributions can be reconstructed using a bivariate Edgeworth series, presented here up to an arbitrarily high order. The Edgeworth formalism is shown to be well-suited for `local' cubic-order non-Gaussianity parametrized by $\gnl$. We show that a strong scale-dependence in the bias can be produced by $\gnl$ of order $10^5$, consistent with CMB constraints. On correlation length of $\sim100$ Mpc, current constraints on $\gnl$ still allow the bias for the most massive clusters to be enhanced by $20-30\%$ of the Gaussian value. We further examine the bias as a function of mass scale, and also explore the relationship between the clustering and the abundance of massive clusters in the presence of $\gnl$. We explain why the Edgeworth formalism, though technically challenging, is a very powerful technique for constraining high-order non-Gaussianity with large-scale structures.

\end{abstract}


\maketitle

\section{Introduction}

One of the most intriguing unanswered questions in cosmology is whether or not the primordial seeds that grew into large-scale structures observed today were laid down as a Gaussian random field. In the simplest single-field inflation model of the early Universe, the initial distribution of the primordial seeds, or density fluctuations, is expected to be very close to Gaussian \cite{bartolo,chen}, but deviations from Gaussianity may be large in more complex models involving multiple fields \cite{rigopoulos,byrnes,chen2,langlois,bartolo2,sasaki} or a non-canonical Lagrangian \cite{alishahiha,arkani-hamed,chen3}. Therefore, a detection of a significant level of primordial non-Gaussianity is of great importance as it would effectively rule out a large class of single-field inflation and open an observational window to the early Universe.

The observational signatures of primordial non-Gaussianity manifest across a large range of physical scales. On very large scales of order several gigaparsecs, non-Gaussianity can be detected, for instance, in the 3-point correlation function (bispectrum) of the cosmic microwave background (CMB) anisotropies (see \cite{bartoloreview,komatsu2} for recent reviews). In the simplest setting in which the bispectrum is parametrized by the constant $\fnl$, the prospect of constraining non-Gaussianity with the CMB seems very  promising indeed. The Planck satellite\footnote{\texttt{http://planck.cf.ac.uk}} will most likely tighten the constraint on $\fnl$ to $\mc{O}$(a few). On smaller scales, the distribution of galaxy clusters can provide competitive constraints on non-Gaussianity, which changes the abundances and clustering properties of large-scale structures (see \cite{desjacquesreview,verde} and references therein). 

A particularly interesting large-scale-structure probe of non-Gaussianity was presented in the seminal work of Dalal \etal \cite{dalal}, who showed quantitatively that non-Gaussianity induces characteristic changes the clustering of density peaks corresponding to rare objects. Specifically, for a correlation length $r$, we can write
\be \xi\sub{pk}(r) = b_L^2(r)\ff \xi(r),\ee
where $\xi\sub{pk}$ denotes the correlation function of density peaks, $\xi$ is that of the underlying dark-matter distribution and $b_L$ is the \ii{bias} parameter (these parameters will be explained in detail later). Physically, the bias quantifies how the density peaks traces of the underlying matter distribution. If the density fluctuations are Gaussian distributed, it can be shown that the bias is almost constant (\ii{i.e.} scale-independent) to a good approximation \cite{kaiser}. The scale-\ii{dependence} of the bias induced by non-Gaussianity is the focus of this work.

Scale-dependent bias from non-Gaussianity is a relatively young but rapidly developing topic. Whilst the dependence of the bias on $\fnl$ was investigated  in \cite{dalal}, a number of authors have since examined the bias for higher-order non-Gaussianity \cite{desjacques}, non-local models \cite{schmidt} and, more recently, scale-dependent $\fnl$ \cite{shandera2} amongst others. The focus of previous works in this area has been the calculation of the bias in Fourier space whilst relying on either numerical simulations or some well-known mass functions. In this work, we show that it is possible to calculate the bias directly in real space by comparing the one- and two-point probability distribution functions (pdfs).

We propose to reconstruct the pdfs by using the Edgeworth series in one and two variables (see \cite{blinnikov,kotz} for reviews). The Edgeworth formalism is a mathematically powerful way to capture the statistical essence of non-Gaussian distributions. In previous astrophysical applications, the Edgeworth series were invariably heavily truncated \cite{scherrer,bernardeau,juszkiewicz,amendola,loverde} yielding pdfs that may not be well-defined, non-negative distributions. In this work, we give a general algorithm which allows the Edgeworth series to be kept to arbitrarily high order. 

We shall see later that given a limited amount of statistical information on the density fluctuations, the Edgeworth formalism is particularly well suited for the reconstruction of non-Gaussian distributions in which the cubic-order non-Gaussianity parameter, $\gnl$, is non-zero. This parameter will be the main focus of our calculations. Once well-defined pdfs are reconstructed, the information on the non-Gaussian bias can then be easily extracted from the one- and two-dimensional pdfs.

\section{The primordial density fluctuations}

We begin by introducing the necessary parameters which will allow us to describe the density fluctuations statistically.

Let $\rho_c$, $\rho_b$, $\rho_r$, $\rho_\Lambda$ be the time-dependent energy densities of cold dark matter, baryons, radiation and dark energy. Let $\rho_m=\rho_c+\rho_b$. We define the density parameter for species $i$ as 
\be \Omega_i \equiv {\rho_i (z=0)\over \rho\sub{crit}},\ee
where $\rho\sub{crit}$ is the critical density defined by $\rho\sub{crit}\equiv 3H_0^2/8\pi G $. The Hubble constant, $H_0$, is parametrized by the usual formula $H_0\equiv100h \mbox{ km\ff s}^{-1}\mbox{Mpc}^{-1}$. Results from a range of astrophysical observations are consistent with $h\simeq 0.7$, $\Omega_c\simeq0.23$, $\Omega_b\simeq0.046$ and $\Omega_r\simeq 8.6\times10^{-5}$, with $\Omega_\Lambda=1-\Omega_m-\Omega_r$ (see \eg \cite{komatsu,lahav+}). 

The density fluctuation field, $\delta$, is defined at redshift $z$ as
\be \delta(\mb{x},z)\equiv {\rho_m(\mb{x},z)-\bkta{\rho_m(z)}\over\bkta{\rho_m(z)}},\ee
where $\bkta{\rho_m}$ is the mean matter energy density. As we are mainly interested in the present-day value of $\delta$, we shall drop the $z$-dependence in our notation and take $\delta=\delta(z=0)$. The Fourier decomposition of $\delta(\mb{x})$ is given by
\be  \delta(\mb{x})=\int {d{\mb k}\over (2\pi)^3} \ff \delta(\mb{k}) e^{i\mb{k}\cdot\mb{x}}.\ee 

The gravitational Newtonian potential $\Phi$ is related to the density fluctuation by the cosmological Poisson equation
\be \delta(\mb{k}) =  {2\over3\Omega_m }\bkt{k\over H_0}^2 \Phi(\mb{k}).\ee
Statistical information on $\delta(\mb{x})$ can be deduced from that of $\delta(\mb{k})$. However, due to the finite resolution of any observation, we can only empirically obtain information on the \ii{smoothed} density field. Given a length scale $R$, the smoothed density field, $\delta_R$, is given by 
\be \delta_R(\mb{k},z)=W(kR)T(k)\delta(\mb{k}),\ee
where $k=|\mb{k}|$. 
We choose $W$ to be the spherical top-hat function of radius $R$. In Fourier space, we have 
\be W(kR)=3\bkts{{\sin(kR)\over (kR)^3}-{\cos(kR)\over (kR)^2}}.\ee
It is also useful to define the mass of matter enclosed by the top-hat window as 
\be M\equiv {4\over3}\pi R^3\rho_m\approx 1.16\times10^{12}\bkt{{R\over h^{-1}\mbox{Mpc}}}^3 \ff h^{-1}M_\sun.\ee
We follow the approach outlined in \cite{weinberg} and use the transfer function $T$ of Dicus 
\begin{align} 
T(x)={\ln[1+(0.124x)^2]\over (0.124x)^2}\bkts{1+(1.257x)^2+(0.4452x)^4+(0.2197x)^6 \over1+(1.606x)^2+(0.8568x)^4+(0.3927x)^6  }^{1/2}.
\end{align}
In addition, we also incorporate the baryonic correction of Eisenstein and Hu \cite{eisenstein}, whereby the transfer function is evaluated at 
\be x\sub{EH}={k\Omega_r^{1/2}\over H_0\Omega_m}\bkts{\alpha+{{1-\alpha}\over{1+(0.43ks)^4}}}^{-1},\lab{EH}\ee
with 
$$\alpha=1-0.328\ln(431\Omega_mh^2){\Omega_b\over\Omega_m}+0.38\ln(22.3\Omega_mh^2)\bkt{\Omega_b\over\Omega_m}^2,$$
and 
$$s={44.5\ln(9.83/\Omega_m h^2) \over \sqrt{1+10(\Omega_b h^2)^{3/4}}} \ff \mbox{Mpc}.$$


The matter power spectrum, $P(k)$, can be defined via the two-point correlation function in Fourier space as
\be \bkta{\delta(\mb{k_1}),\delta(\mb{k_2})}=(2\pi)^3\delta_D(\mb{k_1}+\mb{k_2})P(k),\ee
where $\delta_D$ is the 3-dimensional Dirac delta function. In linear perturbation theory, it is usually assumed that inflation laid down an initial  spectrum of the form $k^{n_s}$, where $n_s$ is the scalar spectral index (assumed to be 0.96 in this work). Physical processes which evolve $P(k)$ through the various cosmological epochs can simply be condensed into the equation
\be P(k)\propto P_\phi(k)T^2(k),\ee
where $P_\phi(k)\propto k^{n_s-4}$. It is also common to define the dimensionless power spectrum $\mc{P}(k)$ as
\be \mc{P}(k)\equiv {k^3\over 2\pi^2} P_\phi(k)\propto \bkt{k\over H_0}^{n_s-1}.\ee
\no Consequently, the variance of density fluctuations smoothed on scale $R$ can be written as 
\be \sigma^2_R(z)= \int_0^\infty {dk\over k} \ff A^2(k,z) \mc{P}(k),\lab{vari}\ee 
where 
\be A(k,z)= {2\over3\Omega_m }\bkt{k\over H_0}^2 T(x\sub{EH}) W(kR).\ee
In our numerical work, we shall normalise $\mc{P}(k)$ so that 
\be \sigma_8\equiv\sigma(R=8h^{-1}\mbox{Mpc},z=0)=0.8.\lab{sig8}\ee

Finally, the correlation function $\xi$ is defined in real space as $\xi(\mb{x_1},\mb{x_2})=\bkta{\delta(\mb{x_1}),\delta(\mb{x_2})}$. If $|\mb{x_1}-\mb{x_2}|=r$, we can write
\be  \xi(r)=\int_0^\infty {dk\over k} \ff A^2(k,z) \mc{P}(k) j_0(kr), \ee
where $j_0(x)=\sin x/x$ (see \eg \cite{lythliddle}). In the limit that $r\rightarrow0$, we recover the auto-correlation \re{vari}.

\section{The clustering of density peaks}\lab{biassection}

The idea that the clustering of density peaks could be measured can be traced back to the pioneering work of Kaiser \cite{kaiser}. Let $P_1$ be the probability that the overdensity at a randomly selected point is above some threshold $\delta_c$, so that
\be P_1 =\int_{\delta_c}^{\infty}p(x)dx,\ee
where $p(x)$ is the pdf for the overdensity. We shall take $p(x)$ to be a weakly non-Gaussian distribution, which permits a valid Edgeworth expansion. This will be discussed in detail in the next section. We take $\delta_c=1.686$, corresponding to the threshold overdensity for spherical collapse.  

Density peaks tend to cluster, and therefore the occurrences of two density peaks are not independent random events. Indeed, the probability that the overdensities at two randomly selected points, separated by comoving distance $r$, both exceed $\delta_c$ is given by 
\be P_2 =\int_{\delta_c}^{\infty}\int_{\delta_c}^{\infty}p(x_1,x_2)dx_1dx_2,\lab{pee2}\ee
where $p(x_1,x_2)$ is the joint pdf. The density-peak correlation function $\xi\sub{pk}$ can be defined as
\be \xi\sub{pk}(r) = {P_2\over P_1^2}-1.\lab{xipeak}\ee
Note that $\xi\sub{pk}=0$ if any two density peaks occur independently. 

The bias parameter, $b_L$, in Lagrangian coordinates is defined as the ratio 
\be b_L^2 = {\xi\sub{pk}(r)\over\xi(r)},\ee
which quantifies the amplitude at which density peaks trace the underlying matter distribution. At late time, what is observable is the Eulerian bias, $b$, 
\be b= 1+b_L.\lab{eulerb}\ee 
If the underlying distribution of $\delta$ were Gaussian, it is well known that in the limit $\delta_c/\sigma_R\gg1$ \cite{kaiser} \be b\sub{Gaussian}\approx 1+{\delta_c\over\sigma^2_R}\ee 
which is scale-independent to a good approximation. Our goal is to quantify the variation in $b$ induced by non-Gaussianity.









\section{The Edgeworth Series}

Equation \re{xipeak} shows that it is possible to calculate the bias directly  once the probability distribution $p(\delta)$ and the joint distribution $p(\delta_1,\delta_2)$ are known. In this section, we shall explain how these distributions can be reconstructed from a few lowest-order moments of the distribution. This technique involves the Edgeworth series, which has been explored by previous authors in simpler forms \cite{scherrer,bernardeau,juszkiewicz,amendola,loverde,me8}. The Edgeworth series  can be summarised schematically as 
\be \mbox{Non-Gaussian pdf} = \mbox{Gaussian}\times (1+ \mbox{deviation}),\ee
where the deviation comprises all known moments of the distribution. In what follows, we define the normalised overdensity as 
\be \nu = {\delta_R\over \sigma_R}, \ee
so that $\bkta{\nu}=1$.

\subsection{Univariate series}

We shall use the form of the univariate Edgeworth series given by Petrov \cite{petrov}, who developed a method for calculating the series to arbitrarily high order. Given a non-Gaussian pdf with zero mean and variance $\sigma_R^2$, we can express its deviation from Gaussianity as a power series in $\sigma_R$:
\be p(\nu) = N(\nu)\bkts{1+\sum_{s=1}^{\infty}\sigma_R^sE_s(\nu)},\lab{edgeworth}\ee
where $N(\nu)$ is the normal distribution
\be N(\nu)={1\over \sigma_R\sqrt{2\pi}}\exp\bkt{- {\nu^2\over 2}},\ee
and the coefficients $E_s$ in the power series are given by
\be
\!\!E_s(\nu)=\!\sum_{\{k_m\}}\bkts{H_{s+2r}(\nu)\prod_{m=1}^{s}{1\over k_m!}\bkt{S_{m+2}\over (m+2)!}^{k_m}\!}.\ff
\lab{coeff}\ee We now explain the various components of the coefficient \re{coeff}. Firstly, the sum is taken over all distinct sets of non-negative integers $\{k_m\}_{m=1}^{s}$ satisfying the Diophantine equation
\be k_1+2k_2+\ldots +sk_s = s.\ee We also define
\be r\equiv k_1+k_2+\ldots +k_s.\ee
Secondly, the function $H_n(\nu)$ is the Hermite polynomial of degree $n$. They can be defined by the Rodrigues' formula
\be H_n(\nu)= (-1)^n e^{\nu^2/2}\diff{^n}{\nu^n}\bkt{e^{-{\nu^2/2}}}.\lab{rod}\ee For example, $H_0(\nu)=1$ and $H_1(\nu)=\nu$. Higher order polynomials can be easily obtained via the recurrence relation
\be H_{n+1}(\nu)=\nu H_n(\nu)-n H_{n-1}(\nu).\lab{recur1}\ee
Thirdly, the reduced cumulants, $S_n$, is defined by
\be S_{n}(R) \equiv  {\bkta{\delta_R^n}_c\over \sigma_R^{2n-2}},\lab{redcum}\ee
where $\bkta{\delta_R^n}_c$ is the $n$th cumulant. For a distribution with zero mean, the relationships between the first few cumulants and moments are
\be \bkta{\delta_R}_c &=& 0, \qquad \bkta{\delta_R^2}_c= \sigma_R^2, \nn\\
\bkta{\delta_R^3}_c &=& \bkta{\delta_R^3}, \quad \bkta{\delta_R^4}_c= \bkta{\delta_R^4}-3\sigma_R^4.\lab{cum}\ee
Note that if $p(\nu)$ is Gaussian, the cumulants of order $\geq3$  vanish identically, and so do the expansion coefficients \re{coeff}, as one might expect. 

Throughout this work we shall often make references to the \ii{skewness} and \ii{kurtosis}, which are defined respectively as $\bkta{\delta_R^3}/\sigma_R^3$ and $\bkta{\delta_R^4}/\sigma_R^4$. The \ii{excess kurtosis} is defined as  as $\bkta{\delta_R^4}/\sigma_R^4-3$, with 3 being the kurtosis of the Gaussian distribution.

\subsection{Bivariate series}

The bivariate Edgeworth series appeared in astrophysical contexts in \cite{lokas,contaldi,lam}, although in these works the series was truncated at low order and resembles a bivariate Gram-Charlier series (see \cite{blinnikov} for detail of  the distinction). In \cite{kota} and \cite{kota2}, the authors presented a bivariate Edgeworth series expanded to an arbitrary number of terms. In this form, the series is given by
\be p(\nu,\nu^\pr) = N(\nu,\nu^\pr)\bkts{1+\sum_{s=1}^{\infty}\sum_{\{P_m\}} \sum_{\{p_i,q_i,\pi_i \}} F\left(\nu,\nu^\pr\right)},\lab{biedgeworth}\ee
where $\nu$ and $\nu^\pr$ are normalised overdensities smoothed on the same scale. The bivariate Gaussian distribution $N(\nu,\nu^\pr)$ is given by
\be N(\nu,\nu^\pr)  = {1\over 2\pi\sqrt{1-\rho^2}}\exp\bkt{-{\nu^2-2\rho \nu\nu^\pr+\nu^{\pr2}\over 2(1-\rho^2)}},\ee
where $\rho$ is the normalized correlation
\be \rho(r)\equiv {\xi(r)\over\sigma^2_R}. \ee
Given an integer $s$, the second sum in \re{biedgeworth} is taken over all distinct sets of positive integers $\{P_m\}_{m=1}^{\ell}$ satisfying the \ii{partition} conditions
\be P_1+P_2+\ldots +P_\ell = s, \lab{par}\\
P_1\geq P_2 \geq \ldots \geq P_\ell>0.\nn\ee

For a given partition $\{P_m\}_{m=1}^{\ell}$, the third sum is taken over all distinct sets of non-negative integers $(p_i,q_i)$ satisfying the \ii{bipartition} condition
\be p_i+q_i=P_i+2.\lab{bipar}\ee
If $(p_i,q_i)$ appears $\pi_i$ times in the bipartition, we write 
\be [P_1P_2\ldots P_\ell]=[(p_1,q_1)^{\pi_1}(p_2,q_2)^{\pi_2}\ldots(p_J,q_J)^{\pi_J}] \quad \mbox{with }\quad\sum_{i=1}^{J}\pi_i=\ell.\ee
As an example, the partitions and bipartitions for the integer $3$ are given in Table \ref{partitiontable}. The number of partitions and bipartitions for integers up to 6 are shown in Table \ref{bitable}.

\begin{table}
\caption{The partitions and bipartitions for the integer $3$.}
\centering
\begin{tabular}{|c|l|}
\hline
Partition [eq. \re{par}]& Bipartition [eq. \re{bipar}]\\
\hline

$[3]$ & $(50),(41),(32),(23),(14),(05)$\\
\hline
$[21]$ & (40)(30),(40)(21),(40)(12),(40)(03),\\
   & (31)(30),(31)(21),(31)(12),(31)(03),\\
   & (22)(30),(22)(21),(22)(12),(22)(03),\\
   & (13)(30),(13)(21),(13)(12),(13)(03),\\
   & (04)(30),(04)(21),(04)(12),(04)(03)\\
   \hline
$[111]$ & $(30)^3,(21)^3,(12)^3,(03)^3,$\\
        & $(30)^2(21),(30)^2(12),(30)^2(03),$\\
        & $(21)^2(30),(21)^2(12),(21)^2(03),$\\
        & $(12)^2(30),(12)^2(21),(12)^2(03),$\\
        & $(03)^2(30),(03)^2(21),(03)^2(12),$\\
        & $(30)(21)(12), (30)(21)(03),(30)(12)(03),(21)(12)(03)$\\
\hline
\end{tabular}
\lab{partitiontable}
\end{table}


\begin{table}
\caption{The number of partitions and bipartitions for some integers.}
\centering
\begin{tabular}{|c|c|c|}
\hline
Integer & \#partitions & \#bipartitions\\
\hline
1 & 1 & 4 \\
2 & 2 & 15 \\
3 & 3 & 46 \\
4 & 5 & 131 \\
5 & 7 & 342 \\
6 & 11 & 851 \\
\hline
\end{tabular}
\lab{bitable}
\end{table}


For each unique bipartition, the function $F$ is given by 
\be F\left( \nu, \nu^\pr\right) &=& \left\{\prod_{i=1}^{J} {1\over \pi_i!}\bkt{\Lambda_{p_iq_i}\over{p_i! q_i!}}^{\pi_i}\right\}H_{\mathsf{p,q}}\bkt{\nu,\nu^\pr},\\
\mathsf{p}=\sum_{i=1}^J p_i\pi_i,&&\mathsf{q}=\sum_{i=1}^J q_i\pi_i.\nn
\ee
Here $H_\mathsf{p,q}$ denotes the bivariate Hermite polynomial defined analogous to \re{rod} as
\be H_\mathsf{p,q}(\nu,\nu^\pr)={(-1)^{\mathsf{p}+\mathsf{q}}\over \widetilde{N}(\nu,\nu^\pr)}{\partial^{\mathsf{p}+\mathsf{q}}\over \partial \nu^\mathsf{p}\partial {\nu^\pr}^\mathsf{q}}\widetilde{N}(\nu,\nu^\pr),\lab{rod2}\\
 \widetilde{N}(\nu,\nu^\pr)\equiv  \exp\bkt{-{\nu^2-2\rho \nu\nu^\pr+{\nu^\pr}^2\over 2(1-\rho^2)}}.\nn\ee
In the Appendix, we outline how $H_\mathsf{p,q}(\nu,\nu^\pr)$ can be efficiently computed. The coefficient $\Lambda_{p,q}$ is defined as 
\be \Lambda_{p,q}(r) = {\bkta{\delta^p \delta^{\pr q}}_c\over \sigma^{p+q}},\ee
where $\delta^\pr\equiv\delta(\mb{x^\pr})$. In other words, $\Lambda_{p,q}$ is the connected part of the correlation between $\hat\delta^p$ and $\hat\delta^{\pr q}$. We shall refer to $\bkta{\delta^p \delta^{\pr q}}_c$ as a joint cumulant (typically there would be a number of joint cumulants of the same order). Similarly, we speak of a joint skewness in the case $p+q=3$, or a joint kurtosis when $p+q=4$.

Finally, note that $F$ contains information on the cumulants of order 3 and higher. One also easily checks that \re{biedgeworth} reduces to the bivariate Gaussian distribution when $F=0$.

\section{Cumulants and Local Non-Gaussianity}

The previous section established the ingredients necessary for the reconstruction of the non-Gaussian pdfs in one and two variables via the Edgeworth series. It is useful to connect these ingredients (which consist of cumulants of the distributions) to a more familiar measure of non-Gaussianity, for example, the parameters $\fnl$ and $\gnl$. 

The most widely studied type of \ng is the `local' type parametrized, at lowest orders, by $\fnl$ and $\gnl$, which are the coefficients in the Taylor expansion of the non-linear Newtonian potential, $\Phi$, in terms of the linear, Gaussian field, $\phi$,
\be \Phi(\mb{x})= \phi(\mb{x})+\fnl\bkt{\phi^2(\mb{x})-\bkta{\phi^2}}+\gnl\phi^3(\mb{x})+\ldots.\lab{expand}\ee
We adopt the `large-scale-structure' convention in which $\Phi$ is extrapolated to $z=0$. We also take $\fnl$ and $\gnl$ to be constant, although it is conceivable that they may be scale-dependent. In this section, we shall calculate the joint skewness and kurtosis as a function of $\fnl$ and $\gnl$ (see \cite{bernardeau,lokas} for previous treatments of the joint cumulants).

\subsection{Joint skewness}

We loosely take {joint skewness} to mean a family of correlations comprising the following quantities
\be \bkta{\delta^3}_c,\ff\bkta{\delta^{\pr3}}_c,\ff\bkta{\delta^2\delta^\pr}_c,\ff\bkta{\delta\delta^{\pr2}}_c. \lab{list}\ee
The first two quantities are equal to the one-point cumulant $\sigma^4 S_3$. It is worth emphasising the subtle difference between $S_3$ and $\Lambda_{3,0}$ 
\be \Lambda_{3,0}=\sigma S_3.\ee  
The remaining two correlations in \re{list} equal 
\be \sigma^3\Lambda_{1,2}(\mb{r})= 2\fnl \int{d\mb{k}\over (2\pi)^3}\int{d\mb{k^\pr}\over (2\pi)^3} A(k)A(k^\pr)A(|\mb{k}+\mb{k^\pr}|)P_\phi(k)P_\phi(k^\pr)\bkts{1+2{P_\phi(|\mb{k}+\mb{k^\pr}|)\over P_\phi(k)}}e^{i(\mb{k}+\mb{k^\pr})\cdot\mb{r}},\lab{jskew}\ee
where $r=|\mb{r}|=|\mb{x_2}-\mb{x_1}|$ \cite{matarrese}. This expression   cannot be analytically evaluated without significant approximations as was done in \cite{bernardeau,lokas}. In this work, we numerically evaluate the joint cumulants directly by a simple change of coordinates. In \re{jskew}, one can align $\mb{r}$ along the ${z}$-axis and introduce spherical coordinates to arrive at

\be \sigma^3\Lambda_{1,2}(r)&=&{\fnl\over 8\pi^2 }\bkt{\prod_{i=1}^{2}\int_0^\infty{dk_i\over k_i}\ff A(k_i) \mc{P}(k_i) \int_{-1}^1\!\! d\mu_i  \int_{0}^{2\pi}\!\! d\phi_i} \ff A(k_{12})\bkts{1+2{P_\phi(k_{12})\over P_\phi(k_2)}}e^{ir(k_1\mu_1+k_2\mu_2)}, \lab{bigjskew}\\
\mbox{where}\ff\ff\ff\ff k_{12}&\equiv&\bkt{k_1^2+k_2^2+2 k_1k_2 \Theta_{12}}^{1/2},\\
\mbox{and}\ff\ff\ff\ff\ff\ff \Theta_{ij}&\equiv&\bkts{(1-\mu_i^2)(1-\mu_j^2)\cos(\phi_i-\phi_j)+\mu_i\mu_j}^{1/2}.\ee 

Note that we can obtain $\Lambda_{3,0}$ by simply evaluating $\Lambda_{1,2}(0)$.

\subsection{Joint kurtosis}

Joint kurtosis refers to three quantities, namely, $\bkta{\delta^4}_c$, $\bkta{\delta^3\delta^\pr}_c$ and $\bkta{\delta^2\delta^{\pr2}}_c$. Again, it is worth pointing out that 
\be \Lambda_{4,0}=\sigma^2 S_4,\ee
and that $\Lambda_{4,0}$ may be obtained from the other 2-point correlations via the relations
\be \Lambda_{4,0}= \Lambda_{3,1}(0)=\Lambda_{2,2}(0).\ee

A change of coordinates again yields the integral expressions for these correlations, 
\be \sigma^4\Lambda_{3,1}(r)&=& \bkt{\prod_{i=1}^{3}\int_0^\infty{dk_i\over k_i}\ff A(k_i) \mc{P}(k_i) \int_{-1}^1\!\! d\mu_i  \int_{0}^{2\pi}\!\! d\phi_i} \ff A(k_{4})\ff e^{ir(k_1\mu_1+k_2\mu_2+k_3\mu_3)}\times\nn\\
&&{3\over32\pi^3}\bigg\{\gnl \bkts{1+3{P_\phi(k_{4})\over P_\phi(k_3)}}+4\fnl^2\mc{I}\bigg\}, \lab{bigjkurt1}\\
\sigma^4\Lambda_{2,2}(r)&=&  \bkt{\prod_{i=1}^{3}\int_0^\infty{dk_i\over k_i}\ff 
A(k_i) \mc{P}(k_i) \int_{-1}^1\!\! d\mu_i  \int_{0}^{2\pi}\!\! d\phi_i} \ff A(k_4)\ff e^{ir(k_1\mu_1+k_2\mu_2)}\times\nn\\
&&{1\over32\pi^3}\bigg\{3\gnl \bkts{1+2{P_\phi(k_{4})\over P_\phi(k_1)}+{P_\phi(k_{4})\over P_\phi(k_3)} }+4\fnl^2\mc{J}\bigg\}, \lab{bigjkurt2}\\
\mbox{where}\ff\ff\ff\ff k_{4}&\equiv&\bkt{k_1^2+k_2^2+k_3^2+2 k_1k_2\Theta_{12}+2k_2k_3\Theta_{23}+2k_1k_3\Theta_{13}}^{1/2}.\ee 
Here $\mc{I}$ and $\mc{J}$ are the contributions of $\fnl$ to the 4-point correlations. The forms of these contributions depend on the symmetries in the integrals above. One can show that
\be\mc{I}={{\mb{(12)}}\over{\mb{(2)}}} \bkt{1+{{\mb{(2)(4)}}\over {\mb{(1)(3)}}}},\ee
\be \mc{J}= {\mb{(13)}\over{\mb{(3)}}}\bkt{1+{{\mb{(3)(4)}}\over {\mb{(1)(2)}}}}+ {{\mb{(12)}+\mb{(23)}}\over{\mb{(2)}}} \bkt{1+{{\mb{(2)(4)}}\over {\mb{(1)(3)}}}},\ee
where we have used the shorthand $\mb{(1)}\equiv P_\phi(k_1)$ and $\mb{(23)}\equiv P_\phi(k_{23})$ \etc\footnote{Setting $\fnl=0$ in equations \re{bigjkurt1}-\re{bigjkurt2}, we recover (A5)-(A6) of \cite{desjacques}. The latter then proceeded with large-scale approximations in Fourier space whereas we have not.}. Since $\mc{I}$ and $\mc{J}$ blow up whenever $k_{12},k_{23}$ or $k_{13}$ vanishes, it is necessary to introduce a large-scale cut-off to evaluate these integrals. To avoid sources of errors associated with this cut-off, we shall only consider the case in which $\fnl=0$. 

\begin{figure}
\vskip -1 in
\centering \hskip 3.5in \includegraphics[width= 11cm, angle = -90]{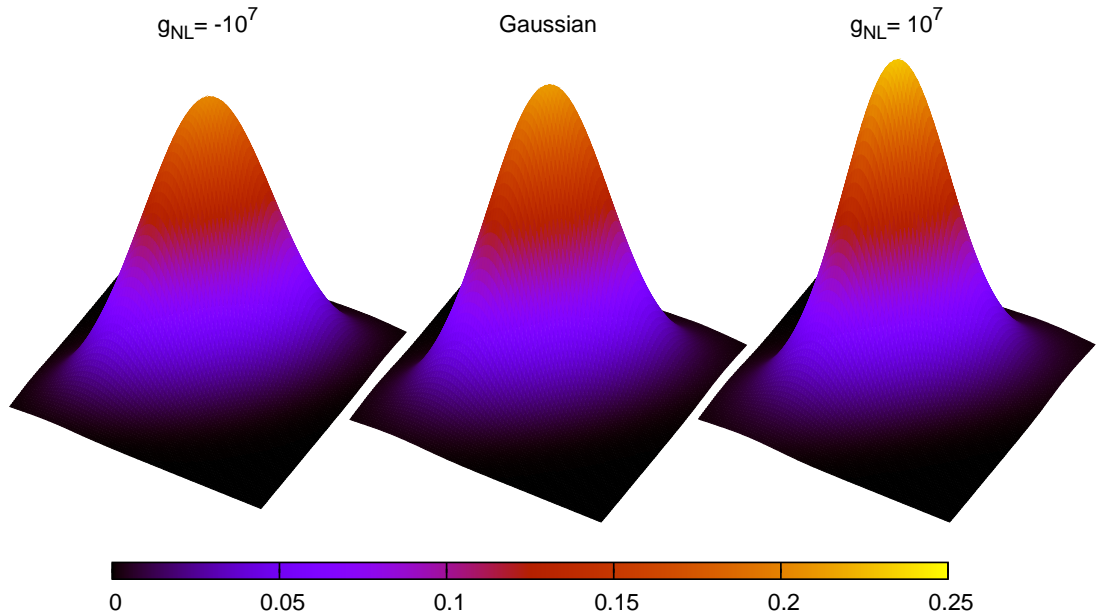}
\caption{The joint pdfs $p(\delta_1,\delta_2)$ corresponding to (left to right) $\gnl=-10^7$, 0 and $10^7$ ($\fnl=0$), reconstructed using the bivariate Edgeworth expansion of order 4, with $R=8h^{-1}$Mpc. The horizontal bar gives the colour code for the probability value on the square grid $[-3,3]^2$. The distributions peak more sharply with increasing $\gnl$. Large values of $\gnl$ have been used for illustrative purposes.}
\label{fig3D}

\end{figure}

Figure \ref{fig3D} shows the joint pdfs with $\gnl=-10^7$, 0 and $10^7$ ($\fnl=0$) reconstructed using the bivariate Edgeworth expansion of order 4. We have chosen large values of $\gnl$ to visually illustrate the effect of $\gnl$ on the joint pdf (namely, the increase in the sharpness of the peak as $\gnl$ increases).


\section{Positivity of the reconstructed pdfs}

Since the reconstructed pdf will be used to calculate the abundance and the  bias of large-scale structures, it is important that the pdf obtained via the Edgeworth series is positive definite. 

\begin{figure}
\includegraphics[width= 6cm, angle = -90]{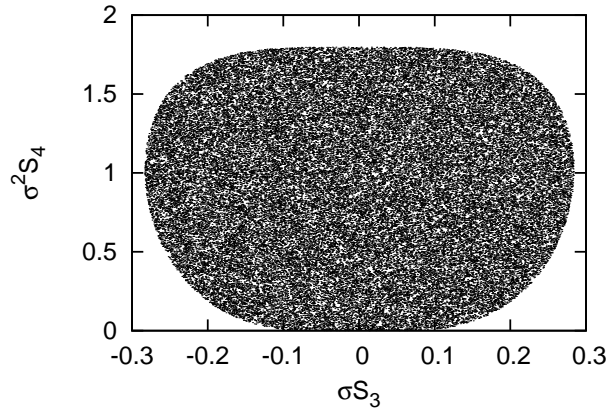}
\caption{ Validity of the 4th-order Edgeworth expansion \re{edgeworth}. The shaded region corresponds to the combinations of $S_3$ and $S_4$ for which there exists a non-negative pdf. On cluster scales where $\sigma\simeq1$, this corresponds to $|\fnl|\lesssim10^3$ and $0\leq\gnl\leq10^8$.}
\label{figscat}

\end{figure}

In general, the positivity of the Edgeworth series is difficult to maintain. As far as we are aware, there exists no general prescriptions that guarantee the positivity of the bivariate Edgeworth series (see \cite{me8} for the analysis of the univariate series). Our investigation shows that the joint pdf tends to develop negative regions whenever the univariate pdf does. For fourth-order series used in this paper, the combinations of $S_3$ and $S_4$ that yield a non-negative pdf are shown in Figure \ref{figscat}. For $\fnl=0$, this corresponds to $\gnl$ in the range $0\lesssim\gnl\lesssim 10^8$. For $\gnl$ outside this range, the reconstructed pdf can develop regions in which $p<0$. This range of validity is well within the 
observational constraints on $\gnl$ (at $2\sigma$):
\be-5.6\times 10^{5}<&\gnl&<6.4\times 10^{5}, \quad \mbox{(Vielva and Sanz \cite{vielva}),}\nn\\
-7.4\times 10^{5}<&\gnl&<8.2\times 10^{5}, \quad \mbox{(Smidt \etals \cite{smidt}),}\nn\\
-3.5\times 10^{5}<&\gnl&<8.2\times 10^{5}, \quad \mbox{(Desjacques and Seljak \cite{desjacques}).}\nn
\ee

In Section \ref{square}, we shall discuss whether it is possible to extend the range of validity of the Edgeworth series to include extreme values of $\gnl$.


\section{Scale-dependent bias induced by $\gnl$}\lab{sectbias}

Using the results in the previous sections, we are now ready to calculate the  bias shift induced by $\gnl$. We summarise the main steps and technical details below.

\bee
\item For a given value of $\gnl$, we calculate the one and two-point cumulants using \re{bigjskew}, \re{bigjkurt1} and \re{bigjkurt2} for a range of values of correlation length $r$. We only consider the case $\fnl=0$ to avoid additional errors from the infrared cut-off in the integrals \re{bigjkurt1}-\re{bigjkurt2}. We initially perform this step at a fixed smoothing scale $R=8h^{-1}$Mpc (the dependence on $R$ will be investigated shortly).
\item The cumulants are then used to reconstruct the univariate and bivariate pdfs using the Edgeworth expansions \re{edgeworth} and \re{biedgeworth} of order $4$. 
\item The reconstructed pdfs are checked to ensure that they are non-negative. For the univariate pdf, this is satisfied when $\gnl$ is in the range $[0,10^8]$. For these values the bivariate pdfs were also found to be non-negative. 
\item Finally, the pdfs are integrated and combined to give the bias $b$ as described in Section \ref{biassection}. 
\eee

\begin{figure}
\centering
\includegraphics[height= 7cm, angle = -90]{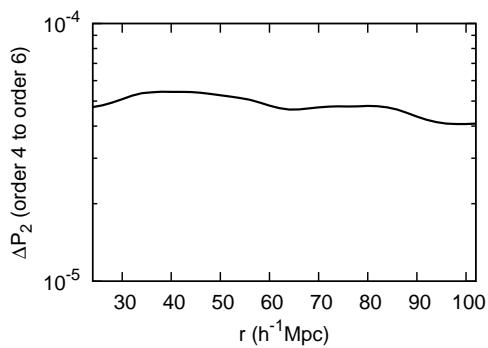}
\caption{ Fractional change in the joint probability $P_2$ (Eq. \ref{pee2}) as the order of the bivariate Edgeworth expansion increases from 4 to 6, plotted against correlation scales (with $\gnl=10^{6}$). The change is less than $0.01$ percent, showing that the expansion \re{biedgeworth} is not highly sensitive to the truncation.}
\label{figspecial}
\end{figure}

It is worth investigating whether the bias is sensitive to the order at which the bivariate series is truncated. First, note that increasing the expansion to fifth-order expansion results in no change in the bivariate pdf (since we have assumed that the odd joint cumulants vanish). Figure \ref{figspecial} shows the fractional change in the joint probability $P_2$ (Eq. \ref{pee2}) expressed as the ratio $|P_2(\mbox{6th order})/P_2(\mbox{4th order}) -1|$ with $\gnl=10^{6}$. We see that the change is less than $0.01$ percent over the range of scales of interest. Thus, we conclude that the bivariate expansion is not strongly sensitive to the truncation order.  This is generally observed for other values of $\gnl$. Considering this modest increase in accuracy at the price of a tremendous increase in the run-time of the code, we find that the 4th order bivariate expansion is adequate for our current investigation\footnote{For detail of the sensitivity of the univariate series to the order of truncation, see \cite{me8}}.


\subsection{Results}


\begin{figure}
\moveright 0.2\linewidth\vbox{
\includegraphics[width= 14cm, angle = -90]{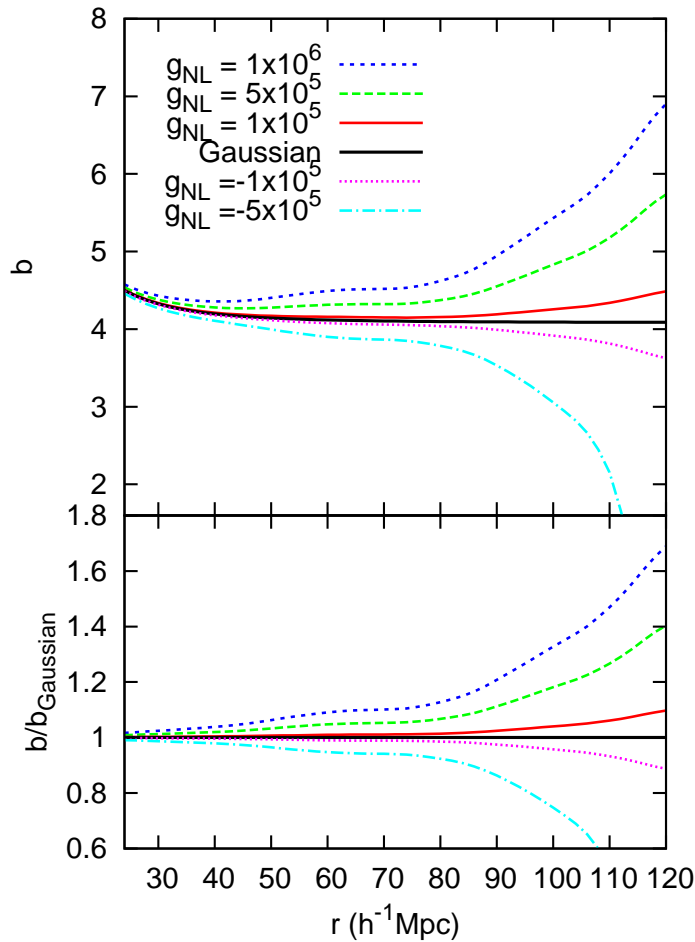}
}
\caption{ The effects of $\gnl$ on the bias $b$ as a function of correlation length $r$. The upper panel shows $b(r)$ for various values of $\gnl$. The lower panel shows the ratio between the non-Gaussian and Gaussian biases. These curves were calculated at smoothing scale $R=8$ $h^{-1}$Mpc, using fourth-order Edgeworth expansions. See the text for more discussion.}
\label{figbias}

\end{figure} 


Figure \ref{figbias} shows the effects of non-Gaussianity on the bias with $\gnl$ up to $10^6$, using the smoothing scale $R=8$ $h^{-1}$Mpc (corresponding to objects of mass $\sim10^{13} h^{-1}M_\sun$ ). The bias is plotted as a function of correlation length of up to $\sim100$ $h^{-1}$Mpc (a typical inter-cluster distance). This is the main result of our work. Note that in the Gaussian case, $b$ is constant on sufficiently large scales to a good approximation.

In general, we observe that large $\gnl$ enhances the clustering of objects on the largest correlation scales. A significant enhancement in the bias can be observed on scales of around 80 Mpc and beyond, consistent with the results of the numerical simulations in \cite{desjacques}. For $\gnl=5\times10^5$ (saturating the CMB constraint) the bias is enhanced by as much as $20-30\%$ at correlation length of $\sim100$ $h^{-1}$Mpc. 

The bias for $\gnl=-10^5$ and  $-5\times10^5$ in \ref{figbias} are included for comparison but should be regarded with caution. As described earlier, the reconstructed pdfs are not positive definite in these cases due to the lack of information on higher-order moments. Nevertheless, we see the general trend that a negative $\gnl$ can significantly suppress the clustering of density peaks.

\subsection{Dependence on the mass scale}\lab{varysect}

\begin{figure}
\moveright 0.15\linewidth\vbox{ \subfloat{\includegraphics[height= 12cm, angle = -90]{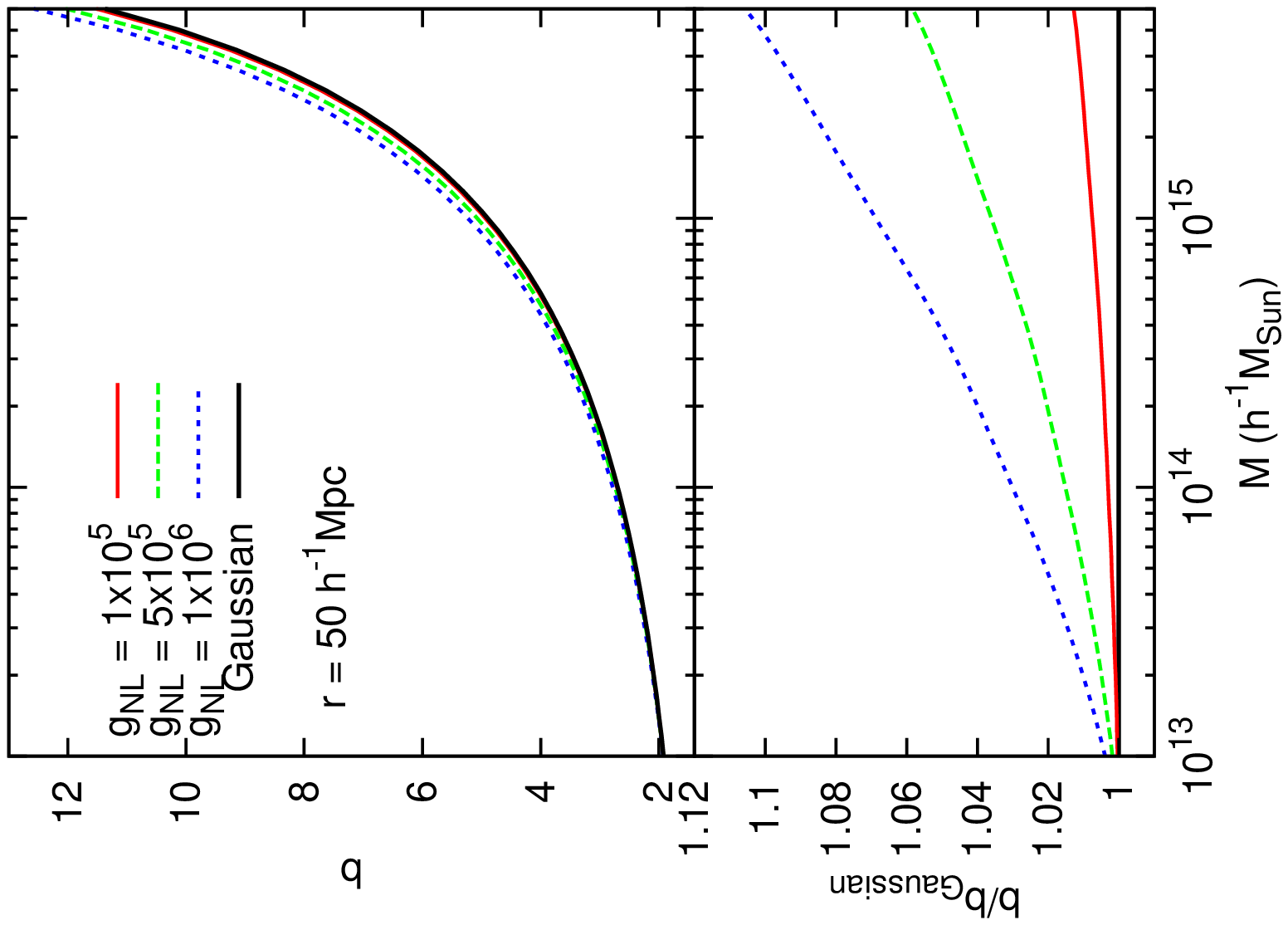}}
\hskip -2.4 in\subfloat{\includegraphics[height= 12cm, angle = -90]{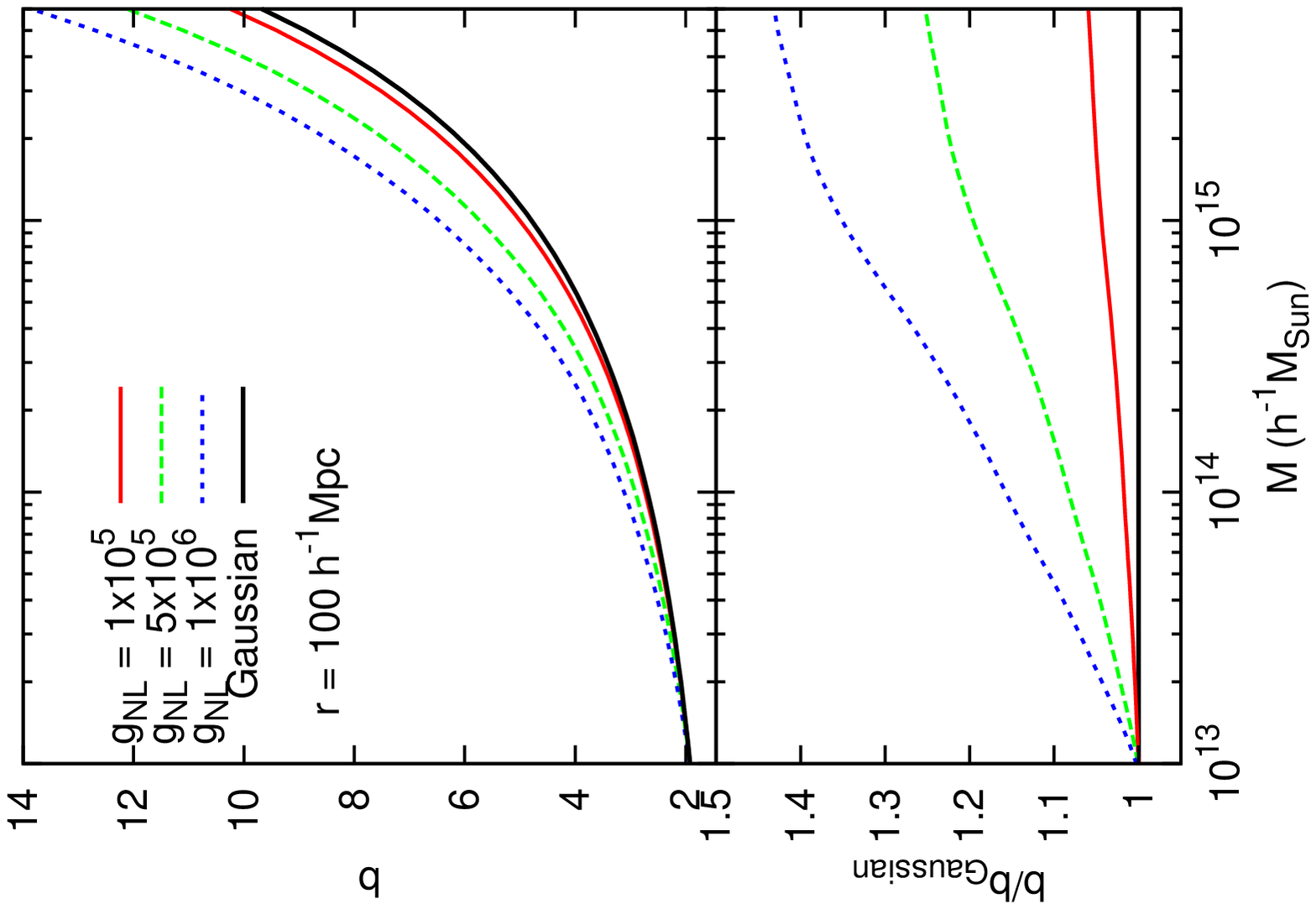}}
}
\begin{flushleft}
\caption{ The effects of $\gnl$ on the bias as a function of the mass scale $M$ for a fixed correlation length $r=50h^{-1}$Mpc (left) and $100h^{-1}$Mpc (right). In each figure, the upper panel shows the bias $b(M)$ for a range of values of $\gnl$, whilst the lower panel shows the ratio of the non-Gaussian and Gaussian biases. See the text for more discussion.}\label{figvary2}
\end{flushleft}
\end{figure}

We now consider the non-Gaussian bias when the smoothing scale $R$, or, equivalently, mass scale $M$, varies while keeping the correlation length fixed. This is useful in determining the effects of non-Gaussianity on the clustering of structures of varying masses for a given correlation length. Figure \ref{figvary2} summarises these effects for $r=50$ and $100$ $h^{-1}$Mpc. In each panel, the bias is plotted as a function of mass scale ($M<10^{16}M_\sun$). In addition, we impose the constraint $r\gtrsim3R$ to avoid nonlinear effects that emerge when the smoothing and correlation scales are comparable.

We observe a monotonic increase in $b$ as $R$ increases, although this dependence is generally weak for a wide range of correlation scales. The monotonic increase in $b/b\sub{Gaussian}$ is observed for smoothing scales $R$ well above the correlation length. The change in curvature seen in the lower panel on the right for $\gnl=10^6$ is most likely a symptom of nonlinear effects as $R\sim r$, and a gradual breakdown of the 4th-order expansion.

At large correlation lengths and in the presence of large $\gnl$, we observe a noticeable enhancement in the bias. For example, at $r\sim100$ Mpc, the bias for the most massive clusters ($M\sim\mbox{a few}\times10^{15}M_\sun$) is enhanced by $20-30\%$ with $\gnl=5\times10^5$. For a shorter correlation length of order a few $\times$ 10  Mpc, $\gnl$ introduces only a sub-percent enhancement in the bias.




\subsection{Clustering versus abundance}

\begin{figure}
\centering
\includegraphics[height= 9.1cm, angle = -90]{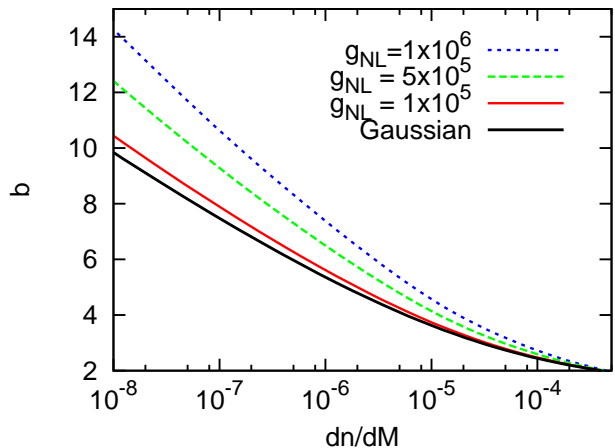}
\caption{ The effect of $\gnl$ on the clustering-abundance relationship for rare objects with correlation length $r=100h^{-1}$Mpc. \ii{Left:} The bias as a function of differential abundance $dn/dM$. Masses of objects in this range vary from $\sim5\times10^{15}M_\sun$ (the rarest clusters) to $10^{13}M_\sun$. }
\label{figversus}
\end{figure}

The two main manifestations of non-Gaussianity in the distribution of large-scale structures are in the abundance and the clustering of rare objects. These effects for $\gnl$ are displayed in Figure \ref{figversus}, which shows the bias as a function of the differential abundance 

\be  \diff{n}{M} = -2{\rho_m\over M}\diff{}{M}\int_{\delta_c /\sigma(M)}^{\infty}p(x,M)dx,\lab{massfun}\ee
where  $n(M)$ is the number density of objects of mass $M$ and $p(x,M)$ is the pdf smoothed by a window function containing mass $M$. On the horizontal axis, the range of masses varies from $10^{16}M_\sun$ (the rarest clusters) down to $10^{13}M_\sun$ (a typical galaxy group). Again, we see the general trend that  both the bias and the abundance of massive clusters increase with $\gnl$ (see \eg \cite{me8} for detailed calculation of the abundance). The non-Gaussian effects are more pronounced for rarer, more massive clusters.


\section{Positivity of the pdf by square-weighting}\lab{square}

Given moments up to order 4 of the distribution of large-scale structures, we have shown that it is possible to construct positive-definite pdfs (in both one and two variables) for $\gnl$ in the range $[0,10^{8}]$. For the technique to be applicable for $\gnl$ outside this range, higher-order moments must be known. A similar conclusion can be drawn for the case of purely $\fnl$-type non-Gaussianity (with $\gnl=0$).

The positivity of the Edgeworth series is a long-standing problem which is not easily overcome. An interesting solution sometimes employed in the economics literature is the square-weighting and renormalisation of the Edgeworth series \citep{gallant,mauleon,perote}. For instance, one could take
\be p(\nu) = {N(\nu)\over C_1}\bkts{1+\sum_{s=1}^{\infty}\sigma_R^sE_s(\nu)}^2,\lab{unipos}\ee
for the univariate series, and similarly,
\be p(\nu,\nu^\pr) = {N(\nu,\nu^\pr)\over C_2}\bkts{1+\sum_{s=1}^{\infty}\sum_{\{P_m\}} \sum_{\{p_i,q_i,\pi_i \}} F\left(\nu,\nu^\pr\right)}^2,\lab{bipos}\ee
for the bivariate series. Here $C_1,C_2$ are constants that renormalise the pdf in each case (note that for the Gaussian case, $C_1=C_2=1$). 

\begin{figure}
\vskip -1.0in
\centering
\hskip 3.6 in \includegraphics[width= 11cm, angle = -90]{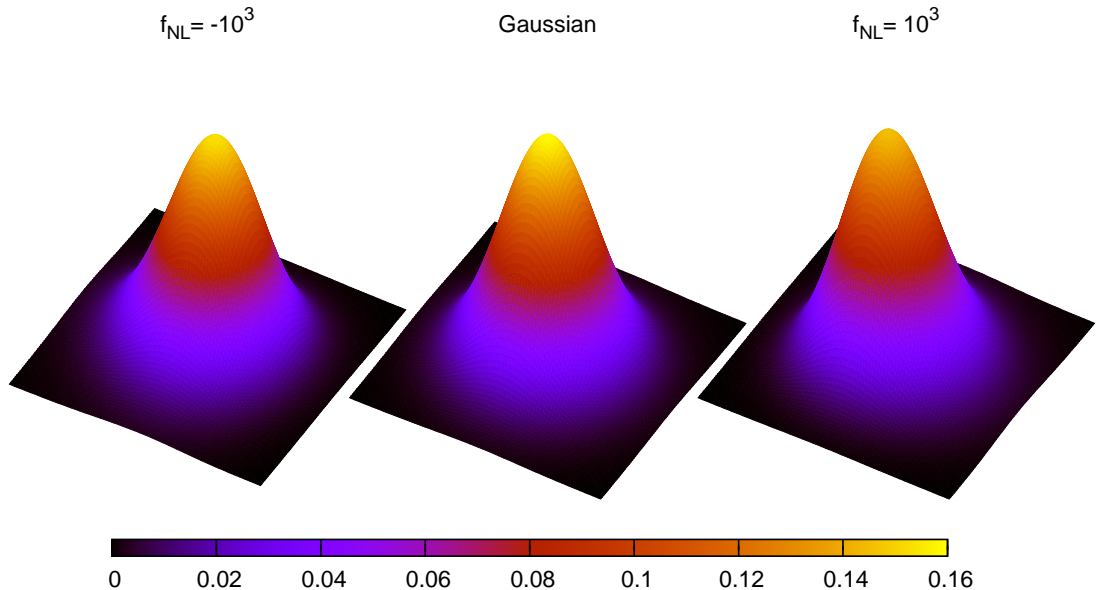}
\caption{The joint pdfs $p(\delta_1,\delta_2)$ corresponding to (left to right) $\fnl=10^{3}$, 0 and $10^{3}$ ($\gnl=0$), reconstructed using the square-weighted bivariate Edgeworth expansion of order 4, with $R=8h^{-1}$Mpc and $r=100h^{-1}$Mpc. The horizontal bar gives the colour code for the probability value on the square grid $[-3,3]^2$. The peak is skewed to the right (towards the first quadrant of the $x-y$ plane) for $\fnl<0$, and left (towards fourth quadrant) for $\fnl>0$.}
\label{fig3Dfnl}

\end{figure}

We have experimented with the square-weighting and found the method to be unsatisfactory. For instance, we found numerical artefacts such as oscillations in the bias due solely to the square-weighting, and are therefore unphysical.  This is not surprising because the square-weighting changes the statistical information of the distribution significantly, and thus the results are difficult to interpret. In addition, there is an order-of-magnitude increase in computing time due to the renormalisation at every time step. Therefore, until further analyses of this sort of square-weighting are performed, we cannot recommend this technique at this point. Nevertheless, for illustrative purposes, we display the reconstructed square-weighted pdf in Figure \ref{fig3Dfnl}, in which large values of positive and negative $\fnl$ skew the pdfs (which are positive definite) in opposite directions as expected.


\section{Conclusions}\lab{conc}

In this work, we have demonstrated an alternative method of calculating the bias in the clustering of rare objects in the presence of primordial non-Gaussianity. Our method is based on the reconstruction of the pdf of density fluctuations using the Edgeworth series in one and two variables. The bias obtained in this way is in real space, in contrast with previous works that examined the scale-dependence bias in Fourier space.

A step-by-step guide to our method is presented in Section \ref{sectbias}. Some of the expressions involved (\eg \re{biedgeworth}) may seem complicated, but this is because they incorporate information on arbitrarily high-order correlations. As long as estimates on these high-order correlations are available, our formalism can, in principle, be used to study the observable signatures of high-order non-Gaussianity. In addition, the reconstruction algorithm is independent of the form of non-Gaussianity, hence making our method easily  applicable to non-local forms of non-Gaussianity as well.

The Edgeworth formalism is a powerful technique that captures all the statistical information of a probability distribution. However, previous astrophysical applications generally dealt with the lowest-order expansions, and therefore the reconstructed pdfs were often found not to be positive definite (in fact, at the lowest order the univariate pdf can never be positive definite). Results obtained from working with pdfs that are not positive definite are unreliable, especially in the context of large-scale structures which are particularly sensitive to the tail end of the pdf.
 
In this work, we concentrate on the case of non-Gaussianity parametrized by positive $\gnl$, which yields pdfs (both uni- and bivariate) that \ii{are} positive definite. It may be surprising to some that the Edgeworth formalism is more easily applied to the case with $\gnl\neq0$ rather than the case with purely $\fnl$-type non-Gaussianity. The reason is that at leading order, $\fnl$ corresponds to the skewness of the distribution. As shown in our previous work \cite{me8}, this information alone cannot define a non-negative pdf. Our previous work also showed that the Edgeworth formalism for the case of pure $\fnl$ requires the knowledge of moments of order at least 5, for which there exist some observational constraints \cite{croton,ross}. The results for $\fnl$ are expected to be similar to that of $\gnl$. This degeneracy can, in theory, be broken by comparing the statistics of voids with that of massive clusters, as any asymmetry in the pdf must be due to the presence of odd-order cumulants. In practice, however, there is the obvious difficulty of determining the abundance and clustering properties of voids. See \cite{damico,me8,lamsheth} for recent progress. 

Our main results show that $\gnl$-type non-Gaussianity can significantly affect the clustering of massive clusters on large correlation scales ($\sim100$ Mpc, typical of inter-cluster distances). A strong scale dependence of the bias can be seen in Figure \re{figbias}, which summarises our main results for $\gnl$ up to $10^6$. It appears that current constraints on $\gnl$ still allow the bias for the most massive clusters to be enhanced by $20-30\%$ of the Gaussian value. Our findings are relevant to observations and $N$-body simulations in which the clustering of extremely massive objects are seen \cite{tian}. An interesting extension of this work is, therefore, a pdf reconstruction using moments observed in large surveys and simulations. It would then be important to include finite-volume effects \cite{kim,bernardeauuzan} which have been shown to systematically alter the cumulants and hence introduce spurious non-Gaussian effects. By using high-order moments and including finite-volume corrections, we expect to be able to extend the Edgeworth formalism to probe a much wider range of high-order non-Gaussianity. This is the subject of our future work.

\mmm
\sss
\bbb

\centerline{\bb{Acknowledgment}}

\mmm

SC is grateful to the referee for many insightful comments. SC supported by Lincoln College, Oxford.

The code (in C++) for generating the bivariate Edgeworth expansion is available upon request.

\bbb

\liner

\appendix

\section{Bivariate Hermite polynomials}

The bivariate Hermite polynomial $H_{n,m}$ is defined via the differential equation \re{rod2}. In this Appendix, we outline a technique which allows $H(n,m)$ to be evaluated efficiently. 

Firstly, we assume $m\geq n$, otherwise one may appeal to the identity
\be H_{n,m}(x,y) = H_{m,n}(y,x),\ee 
which can be deduced from \re{rod2}. The numerical value of $H_{m,n}(x,y)$ can be computed using the recurrence relation first obtained by Hermite himself \citep{hermite}
\be H_{n,m+1}(x,y)={1\over 1-\rho^2}\bkts{(y-\rho x)H_{n,m}(x,y)+\rho n H_{n-1,m}(x,y)-m H_{n,m-1}(x,y)}, \quad n,m\geq1.\ee
This recurrence requires the knowledge of $H_{1,1}$ and $H_{0,m}$ for $m\geq0$. It is straightforward to evaluate $H_{1,1}$ directly from \re{rod2}, giving
\be H_{1,1}(x,y)= {(y-\rho x)(x-\rho y)\over (1-\rho^2)^2} +{\rho\over 1-\rho^2}.\ee
For $H_{0,m}$, a simple change of variable gives
\be H_{0,m}(x,y)= (1-\rho^2)^{-m/2}\ff H_m\bkt{y-\rho x\over \sqrt{1-\rho^2}},\ee
where $H_m$ is the standard Hermite polynomial.

\bibliographystyle{h-physrev3}
\bibliography{ng_bias}

\begin{thebibliography}{10}

\bibitem{bartolo}
N.~Bartolo, E.~Komatsu, S.~Matarrese, and A.~Riotto,
\newblock Phys. Rept. {\bf 402}, 103 (2004).

\bibitem{chen}
X.~Chen,
\newblock (2010), 1002.1416.

\bibitem{rigopoulos}
G.~I. Rigopoulos, E.~P.~S. Shellard, and B.~J.~W. van Tent,
\newblock Phys. Rev. {\bf D73}, 083522 (2006).

\bibitem{byrnes}
C.~T. {Byrnes} and K.~{Choi},
\newblock (2010), 1002.3110.

\bibitem{chen2}
X.~Chen,
\newblock Phys. Rev. {\bf D72}, 123518 (2005).

\bibitem{langlois}
D.~Langlois, S.~Renaux-Petel, D.~A. Steer, and T.~Tanaka,
\newblock Phys. Rev. Lett. {\bf 101}, 061301 (2008).

\bibitem{bartolo2}
N.~Bartolo, S.~Matarrese, and A.~Riotto,
\newblock Phys. Rev. D {\bf 69}, 043503 (2004).

\bibitem{sasaki}
M.~Sasaki, J.~V\"aliviita, and D.~Wands,
\newblock Phys. Rev. D {\bf 74}, 103003 (2006).

\bibitem{alishahiha}
M.~Alishahiha, E.~Silverstein, and D.~Tong,
\newblock Phys. Rev. D {\bf 70}, 123505 (2004).

\bibitem{arkani-hamed}
N.~Arkani-Hamed, P.~Creminelli, S.~Mukohyama, and M.~Zaldarriaga,
\newblock JCAP {\bf 0404}, 001 (2004).

\bibitem{chen3}
X.~Chen, M.-x. Huang, S.~Kachru, and G.~Shiu,
\newblock JCAP {\bf 0701}, 002 (2007).

\bibitem{bartoloreview}
N.~Bartolo, S.~Matarrese, and A.~Riotto,
\newblock (2010), 1001.3957.

\bibitem{komatsu2}
E.~{Komatsu},
\newblock Classical and Quantum Gravity {\bf 27}, 124010 (2010).

\bibitem{desjacquesreview}
V.~{Desjacques} and U.~{Seljak},
\newblock Classical and Quantum Gravity {\bf 27}, 124011 (2010).

\bibitem{verde}
L.~{Verde},
\newblock Advances in Astronomy {\bf 2010} (2010).

\bibitem{dalal}
N.~Dalal, O.~Dore, D.~Huterer, and A.~Shirokov,
\newblock Phys. Rev. {\bf D77}, 123514 (2008).

\bibitem{kaiser}
N.~{Kaiser},
\newblock Astrophys. J. Lett. {\bf 284}, L9 (1984).

\bibitem{desjacques}
V.~Desjacques and U.~Seljak,
\newblock Phys. Rev. {\bf D81}, 023006 (2010).

\bibitem{schmidt}
F.~{Schmidt} and M.~{Kamionkowski},
\newblock ArXiv e-prints  (2010), 1008.0638.

\bibitem{shandera2}
S.~{Shandera}, N.~{Dalal}, and D.~{Huterer},
\newblock ArXiv e-prints  (2010), 1010.3722.

\bibitem{blinnikov}
S.~{Blinnikov} and R.~{Moessner},
\newblock Astron. Astrophys. Supp. {\bf 130}, 193 (1998).

\bibitem{kotz}
S.~Kotz, N.~Balakrishnan, and N.~L. Johnson,
\newblock {\em Continuous Multivariate Distributions, Vol 1: Models and
  Applications,}Wiley Series in Probability and Statistics, 2 ed. (John Wiley
  \& Sons, 2000).

\bibitem{scherrer}
R.~J. {Scherrer} and E.~{Bertschinger},
\newblock \apj {\bf 381}, 349 (1991).

\bibitem{bernardeau}
F.~{Bernardeau},
\newblock Astron. Astrophys. {\bf 312}, 11 (1996).

\bibitem{juszkiewicz}
R.~{Juszkiewicz}, D.~H. {Weinberg}, P.~{Amsterdamski}, M.~{Chodorowski}, and
  F.~{Bouchet},
\newblock \apj {\bf 442}, 39 (1995).

\bibitem{amendola}
L.~{Amendola},
\newblock \apj {\bf 569}, 595 (2002).

\bibitem{loverde}
M.~LoVerde, A.~Miller, S.~Shandera, and L.~Verde,
\newblock JCAP {\bf 0804}, 014 (2008).

\bibitem{komatsu}
E.~Komatsu {\em et~al.},
\newblock (2010), 1001.4538.

\bibitem{lahav+}
O.~Lahav and A.~R. Liddle,
\newblock (2010), 1002.3488.

\bibitem{weinberg}
S.~Weinberg,
\newblock {\em Cosmology} (Oxford University Press, 2008).

\bibitem{eisenstein}
D.~J. Eisenstein and W.~Hu,
\newblock Astrophys. J. {\bf 496}, 605 (1998).

\bibitem{lythliddle}
D.~H. Lyth and A.~R. Liddle,
\newblock {\em The Primordial Density Perturbation} (Cambridge University
  Press, 2009).

\bibitem{me8}
S.~{Chongchitnan} and J.~{Silk},
\newblock \apj {\bf 724}, 285 (2010).

\bibitem{petrov}
V.~Petrov,
\newblock {\em Sums of Independent Random Variables, \ff}volume~82 of {\em
  Ergebnisse der Mathematik und ihrer Grenzgebiete} (Springer-Verlag, Berlin,
  1975).

\bibitem{lokas}
E.~L. Lokas,
\newblock (1997), astro-ph/9708047.

\bibitem{contaldi}
C.~R. {Contaldi}, P.~G. {Ferreira}, J.~{Magueijo}, and K.~M. {G{\'o}rski},
\newblock \apj {\bf 534}, 25 (2000).

\bibitem{lam}
T.~Y. Lam and R.~K. Sheth,
\newblock (2009), 0905.1702.

\bibitem{kota}
V.~K.~B. {Kota},
\newblock Zeitschrift fur Physik A Hadrons and Nuclei {\bf 315}, 91 (1984).

\bibitem{kota2}
V.~K.~B. Kota, K.~B.~K. Mayya, and J.~A.~C. Alcaras,
\newblock Journal of Physics A: Mathematical and Theoretical {\bf 42}, 145201
  (2009).

\bibitem{matarrese}
S.~Matarrese and L.~Verde,
\newblock Astrophys. J. {\bf 677}, L77 (2008).

\bibitem{vielva}
P.~{Vielva} and J.~L. {Sanz},
\newblock Mon. Not. Roy. Astron. Soc. {\bf 404}, 895 (2010).

\bibitem{smidt}
J.~Smidt {\em et~al.},
\newblock (2010), 1001.5026.

\bibitem{gallant}
A.~R. Gallant and D.~W. Nychka,
\newblock Econometrica {\bf 55}, pp. 363 (1987).

\bibitem{mauleon}
I.~Maule\'{o}n and J.~Perote,
\newblock The European Journal of Finance {\bf 6}, 225 (2000).

\bibitem{perote}
J.~Perote and E.~del Br{\'\i}o,
\newblock International Advances in Economic Research {\bf 12}, 425 (2006).

\bibitem{croton}
2dFGRS Team, D.~J. Croton {\em et~al.},
\newblock Mon. Not. Roy. Astron. Soc. {\bf 352}, 1232 (2004).

\bibitem{ross}
A.~J. {Ross}, R.~J. {Brunner}, and A.~D. {Myers},
\newblock \apj {\bf 649}, 48 (2006).

\bibitem{damico}
G.~{D'Amico}, M.~{Musso}, J.~{Nore{\~n}a}, and A.~{Paranjape},
\newblock ArXiv e-prints  (2010), 1011.1229.

\bibitem{lamsheth}
T.~Y. {Lam}, R.~K. {Sheth}, and V.~{Desjacques},
\newblock Mon. Not. Roy. Astron. Soc. {\bf 399}, 1482 (2009).

\bibitem{tian}
H.~J. {Tian}, M.~C. {Neyrinck}, T.~{Budav{\'a}ri}, and A.~S. {Szalay},
\newblock ArXiv e-prints  (2010), 1011.2481.

\bibitem{kim}
R.~S. {Kim} and M.~A. {Strauss},
\newblock \apj {\bf 493}, 39 (1998).

\bibitem{bernardeauuzan}
F.~Bernardeau and J.-P. Uzan,
\newblock Phys. Rev. {\bf D70}, 043533 (2004).

\bibitem{hermite}
C.~Hermite,
\newblock Comptes rendus de l'Acad\'{e}mie des Sciences {\bf 58}, 266 (1864).

\end{thebibliography}

\end{document}